\documentstyle[pra,aps,eqsecnum]{revtex}
\begin{document}
\def\bx{\mbox{\boldmath $\xi$}}
\def\bg{\mbox{\boldmath $\gamma$}}
\def\bb{\mbox{\boldmath $\beta$}}
\def\ba{\mbox{\boldmath $\alpha$}}
\def\bs{\mbox{\boldmath $\psi$}}
\title{Parametrizing the mixing matrix : A unified approach} 
\author{S. Chaturvedi\thanks{e-mail: scsp@uohyd.ernet.in}}
\address{School of Physics, University of Hyderabad, Hyderabad 500 046, India}
\author{N. Mukunda\thanks{email: nmukunda@cts.iisc.ernet.in}}
\address{Centre for Theoretical Studies and Department of Physics, 
Indian Institute of Science, Bangalore 560 012, India, and\\
Jawaharlal Nehru Centre for Advanced Scientific Research, 
Jakkur, Bangalore 560 064, India}
\date{\today}
\maketitle
\begin{abstract}
A unified approach to parametrization of the mixing matrix for $N$ 
generations is developed. This approach not only has a clear geometrical 
underpinning but also has the advantage of being economical and recursive 
and leads in a natural way to the known phenomenologically useful 
parametrizations of the mixing matrix.
\end{abstract}
\section{Introduction}
In the standard $SU(3)\times SU(2)\times U(1)$ model of strong , weak and 
electromagnetic interactions, all aspects of the charged weak interactions 
among quarks can be described in terms of a $3\times 3$ unitary matrix 
\begin{equation}
V = \left(\begin{array}{ccc} V_{ud}&V_{us}&V_{ub}\\ 
V_{cd}&V_{cs}&V_{cb}\\
V_{td}&V_{ts}&V_{tb}
\end{array}\right)
\end{equation}
specified by four real parameters: three generalized Cabibbo angles and one 
Kobayashi-Maskawa phase. After the pioneering work of Kobayashi and 
Maskawa [1], this matrix, which describes the mixing between quark mass 
eigenstates and the charged weak current eigenstates, has been parametrized 
in a number of phenomenologically useful ways [2-6]. Generalizations to 
$N\geq 3$ generations of quarks, where the mixing matrix is characterized by 
$N(N-1)/2$ angles and $(N-1)(N-2)/2$ phases, have also been proposed [6-14]. 
Analogues of the mixing matrix also arise in the lepton sector if the 
neutrinos are taken as massive Dirac particles. In most of the 
parametrizations hitherto proposed, the mixing matrix is expressed as an 
ordered product of $N(N-1)/2$ factors each of which carries an angle. Of 
these $N(N-1)/2$ factors, a prescribed set of $(N-1)(N-2)/2$ factors 
carry phases as well. Different parametrizations differ from each other in 
the ordering prescription and the location of the phase factors within the 
matrices carrying them. In this work, we present a parametrization of the 
mixing matrix based on a decomposition,  involving, in the $N=3$ case, just 
two factors. This 
parametrization, apart from having a clear geometrical picture underlying 
it, also enables us to recover and relate other parametrizations and to 
generate new ones in a unified manner.

\section{Parametrizing SU(n) elements by a sequence of complex unit vectors} 
The proposed parametrization of the mixing matrix is based on the observation 
that a generic matrix $g\in SU(N)$ can be parametrized by a sequence of 
complex unit vectors $\bx,\cdots,\bg,\bb,\ba$ of dimensions $n,n-1, \cdots,3,2$
. This can be seen as follows. Let 
\begin{equation}
\Sigma_n = \{ \bs=\left(\begin{array}{c}\psi_1\\ \psi_2\\ \cdot \\ \cdot \\ 
\psi_n \end{array}\right) \mid \bs^\dagger\bs=1\}
\end{equation}
denote the set of unit vectors in complex n-dimensional Hilbert space i.e. 
a set of real dimension $(2n-1)$. Any $\bs\in \Sigma_n$ can be mapped to 
the vector 
\begin{equation}
{\bf e}_n=\left(\begin{array}{c}0\\ 0\\ \cdot \\ \cdot \\ 
1 \end{array}\right)
\end{equation}
via a suitable $SU(n)$ element. ( Note that we are really using $SU(n)$ here ,
not $U(n)$)   Therefore, $SU(n)$ acts transitively on $\Sigma_n$. The subgroup 
of $SU(n)$ that leaves ${\bf e}_n$ invariant is $SU(n-1)$ on the first $(n-1)$ 
dimensions and hence 
\begin{equation}
\Sigma_n \simeq ~coset~ space~ SU(n)/SU(n-1)
\end{equation} 
Therefore, we expect that, apart from global matching problems or ambiguities 
on a subset of measure zero, any element in  $SU(n)$ is uniquely specified by 
a pair consisting of an element in $SU(n-1)$ and a unit vector $\bs 
\in \Sigma_n$. Therefore, recursively, we see that an element $g\in SU(n)$ 
can be parametrized as 
$g=g(\bx,\cdots,\bg,\bb,\ba)$
by a string of complex unit vectors 
$\bx,\cdots,\bg,\bb,\ba$ of dimensions $n,n-1, \cdots,4,3,2$. 
  
As a convention, we will let the above unit vectors stand for the last column. 
in the relevant $SU(n)$ matrices. This is because when a matrix of $SU(n)$ is 
multiplied on the right by a matrix of $SU(n-1)$ (leaving ${\bf e}_n$ 
invariant), it is the last column in the former matrix that remains unchanged.
For elements of $SU(2)$ we will thus write    
\begin{equation}
A\in SU(2)~~;~~
A = A(\ba)= \left(\begin{array}{cc} \alpha_{2}^{*}&\alpha_1\\ 
-\alpha_{1}^{*}&\alpha_2\\
\end{array}
\right)~~~;~~~\ba^\dagger \ba=1
\end{equation}
This is globally well defined.

\section{SU(3) and the Kobayashi-Maskawa phase}

Let $\bb$ denote a three component complex unit vector, $\bb^\dagger\bb =1$. 
Then for $|\beta_1|~<~1$, the matrix
\begin{equation}
B(\bb)  = \left(\begin{array}{ccc} P^{-1}&0&\beta_1\\ 
-P\beta_1^{*}\beta_{2}&P\beta_{3}^{*}&\beta_2\\
-P\beta_1^{*}\beta_{3}&-P\beta_{2}^{*}&\beta_3
\end{array}
\right)~~,~~ P= (1-|\beta_1|^2)^{-1/2}
\end{equation}
is in $SU(3)$. the unit vector $\bb$ is a label for right $SU(2)$ 
cosets in $SU(3)$, and $B(\bb)$ is a coset representative.  So any 
$B\in SU(3)$, $|B_{13}| <1$, can be uniquely written as 
\begin{equation}
B  = B(\bb,\ba) =B(\bb)\cdot\left(\begin{array}{cc} A(\ba)&0\\ 
0&1\end{array}\right).
\end{equation}
On multiplying out the two matrices on the rhs one obtains
\begin{equation}
B(\bb,\ba)  = \left(\begin{array}{ccccc} P^{-1}\alpha_{2}^{*}&&P^{-1}\alpha_1
&&\beta_1\\ 
-P\beta_1^{*}\beta_{2}\alpha_{2}^{*}-P\beta_{3}^{*}\alpha_{1}^{*}&&
-P\beta_{1}^{*}\beta_2\alpha_1+P\beta_{3}^{*}\alpha_2 &&\beta_2\\
-P\beta_1^{*}\beta_{3}\alpha_{2}^{*}+P\beta_{2}^{*}\alpha_{1}^{*} &&
-P\beta_{1}^{*}\beta_3\alpha_1-P\beta_{2}^{*}\alpha_2&&\beta_3 \end{array}
\right)
\label{1}
\end{equation} 
Now we examine how $B(\bb,\ba)$ transforms under rephasing transformations i.e.
we ask how $\bb$ and $\ba$ change when  we multiply $B(\bb,\ba)$ on the left 
and on the right by independent diagonal elements of $SU(3)$:
\begin{equation}
B^\prime = D(\theta^\prime)~ B~ D(\theta)
\end{equation}
where $D(\theta)=diag(e^{i\theta_1+i\theta_2}, e^{-i\theta_1+i\theta_2},
e^{-2i\theta_2})$ and $D(\theta^\prime)$ is defined similarly. 
Then we find
\begin{equation}
D(\theta^\prime)B(\bb,\ba)D(\theta)= B(\bb^\prime,\ba^\prime)
\label{6}
\end{equation}
where
\begin{eqnarray}
\alpha_{1}^{\prime} &=& \alpha_{1}e^{i(\theta_{1}^\prime+
\theta_{2}^{\prime}-\theta_{1}+\theta_2)}~~~;~~~
\alpha_{2}^{\prime} = \alpha_{2}e^{-i(\theta_{1}^\prime+
\theta_{2}^{\prime}+\theta_{1}+\theta_2)}
\label{7}\\
\beta_{1}^{\prime}& =& \beta_{1}e^{i(\theta_{1}^\prime+\theta_{2}^\prime
-2\theta_2)}~;~
\beta_{2}^{\prime} = \beta_{2}e^{i(-\theta_{1}^\prime+\theta_{2}^\prime
-2\theta_2)}~;~
\beta_{3}^{\prime} = \beta_{3}e^{-2i(\theta_{2}^\prime+\theta_{2})}
\label{8}
\end{eqnarray} 
These transformation laws can easily be written down from the  locations of 
$\alpha_1,\alpha_2,\beta_1, \beta_2, \beta_3$ in the matrix $(\ref{1})$.

As the dimension of $SU(3)$ is eight and we have four independent phases here, 
there should be four independent real invariants. Three of them, essentially 
the generalized Cabibbo angles may be chosen to be, say, 
$|\alpha_1|, |\beta_1|,|\beta_2|$. The fourth one can be found systematically 
as follows. 

The phase $\theta_1$ does not occur in $\bb^\prime$. So we form the 
combination $\alpha_1\alpha_{2}^{*}$ whose transformation law is $\theta_1$ 
independent: 
\begin{equation}
\alpha_{1}^{\prime} \alpha_{2}^{\prime*}=\alpha_1\alpha_{2}^{*}e^{2i(\theta_{1}
^{\prime} +\theta_{2}^{\prime}+\theta_2)}
\label{2}
\end{equation}
Among the $\beta$'s, $\theta_{1}^{\prime}$ occurs only in $\beta_{1}^{\prime}$
and $\beta_{2}^{\prime}$, so we form a combination which can cancel 
$e^{2i\theta_{1}^{\prime}}$ on the rhs of $(\ref{2})$
\begin{equation}
\beta_{1}^{*}\beta_2 \rightarrow \beta_{1}^{*}\beta_2 e^{-2i\theta_{1}^\prime}
\label{3}
\end{equation}
From $(\ref{2})$ and $(\ref{3})$ we find 
\begin{equation}
\alpha_1\alpha_{2}^{*}\beta_{1}^{*}\beta_2 \rightarrow
\alpha_1\alpha_{2}^{*}\beta_{1}^{*}\beta_2 e^{2i(\theta_{2}^\prime  +\theta_2)}
\end{equation}
Comparing this with $\beta_{3}^{\prime}$  we see that
$arg(\alpha_1\alpha_{2}^{*}\beta_{1}^{*}\beta_2 \beta_3)$ is invariant under
rephasing. 

\section{Comparison with some well known parametrizations of the mixing 
matrix for N=3}   

Before we show how some well known  parametrizations of the mixing matrix can 
be recovered from the considerations given above, it is useful to note that 
from the standard form$(\ref{1})$ we can generate others by permutation of 
rows and columns and by taking transpose. The expressions for the invariants 
remain unchanged under these opertations as will become clear in section VII.
This being the case, various parametrizations of the mixing matrix can be 
generated by choosing any one from 
$\alpha_1,\alpha_{2},\beta_{1},\beta_2 ,\beta_3$ which appear in the 
invariant
$arg(\alpha_1\alpha_{2}^{*}\beta_{1}^{*}\beta_2 \beta_3)$ to be complex 
and all others real in the matrix $(\ref{1})$ or in the matrices obtained 
by permuting rows and columns or by taking transpose. Thus, for instance, 
choosing $\beta_2$ to be complex, all others real, and putting   
\begin{equation}      
\alpha_1= S_\theta, \alpha_2= C_\theta, \beta_1 = S_\beta, \beta_2 = 
S_\gamma C_\beta e^{i\delta}, \beta_3 = C_\gamma C_\beta 
\end{equation}
in $(\ref{1})$ one obtains the Maiani parametrization [2]
\begin{equation}
 \left(\begin{array}{ccc} C_\beta C_\theta&C_\beta S_\theta
&S_\beta\\ 
-S_\gamma C_\theta S_\theta e^{i\delta}-S_\theta C_\gamma&
C_\gamma C_\theta - S_\gamma S_\beta S_\theta e^{i\delta}&S_\gamma C_\beta 
e^{i\delta}\\
-S_\beta C_\gamma C_\theta + S_\gamma S_\theta e^{i\delta}&
-C_\gamma S_\beta S_\theta - S_\gamma C_\theta e^{-i\delta}&C_\gamma C_\beta 
\end{array}\right)
\end{equation}

The Chau-Keung parametrization [4] corresponds to choosing $\beta_1$ complex. 
\begin{equation}      
\alpha_1= S_{12}, \alpha_2= C_{12}, \beta_1 = S_{13}e^{-i\delta}, \beta_2 = 
S_{23} C_{13} , \beta_3 = C_{23} C_{13} 
\end{equation}
The mixing matrix, for this choice, is given by 
\begin{equation}
 \left(\begin{array}{ccc} C_{12} C_{13}&S_{12} C_{13}
&S_{13}e^{-i\delta_{13}}\\ 
-S_{12} C_{23}-C_{12} S_{23} S_{13} e^{i\delta_{13}}&
C_{12} C_{23} - S_{12} S_{23} S_{13} e^{i\delta_{13}}&S_{23} C_{13}\\
S_{12}S_{23}- C_{12} C_{23} S_{13} e^{i\delta_{13}}&
-C_{12} S_{23}- S_{12}C_{23}S_{13}e^{i\delta_{13}}&C_{23} C_{13} 
\end{array}\right)
\end{equation}
The Kobayashi-Maskawa form corresponds to taking $\beta_2$ complex and 
putting
\begin{equation}      
\alpha_1= -S_{3}, \alpha_2= -C_{3}, \beta_1 = C_{1}, \beta_2 = 
-S_{1} C_{2}e^{-i\delta} , \beta_3 = S_{1} S_{2} 
\end{equation}
in
\begin{equation}
B(\bb,\ba)  = \left(\begin{array}{ccccc} \beta_1&&P^{-1}\alpha_{2}^{*}&&
P^{-1}\alpha_1
\\ 
\beta_2&&-P\beta_1^{*}\beta_{2}\alpha_{2}^{*}-P\beta_{3}^{*}\alpha_{1}^{*}&&
-P\beta_{1}^{*}\beta_2\alpha_1+P\beta_{3}^{*}\alpha_2 \\
\beta_3&&-P\beta_1^{*}\beta_{3}\alpha_{2}^{*}+P\beta_{2}^{*}\alpha_{1}^{*} &&
-P\beta_{1}^{*}\beta_3\alpha_1-P\beta_{2}^{*}\alpha_2 \end{array}
\right)
\end{equation} 
and leads to 
\begin{equation}
\left(\begin{array}{ccc} C_{1} &-S_{1}C_{3}
&-S_{1}S_{3}\\ 
-S_{1} C_{2}e^{-i\delta}&
-C_{1} C_{2}C_{3}e^{-i\delta} + S_{2} S_{3}&-C_{1}C_{2} S_{3}e^{-i\delta}-
S_{2}C_{3}\\
S_{1}S_{2}&C_{1} S_{2} C_{3}+C_{2}S_{3} e^{i\delta}&
C_{1} S_{2}S_{3}- C_{2}C_{3}e^{i\delta} 
\end{array}\right)
\end{equation}
which, on multiplying the second row by the phase factor $(e^{-i\delta})$ 
gives precisely the mixing matrix originally given by Kobayashi and Maskawa.   

Similarly, taking $\beta_2$ to be complex and putting 

\begin{equation}      
\alpha_1= C_{13}, \alpha_2= -S_{13}, \beta_1 = -S_{12}, \beta_2 = 
C_{12} C_{23}e^{-i\alpha} , \beta_3 = C_{12} S_{23} 
\end{equation}
in 
\begin{equation}
B(\bb,\ba)  = \left(\begin{array}{ccccc} P^{-1}\alpha_1
&&\beta_1&&P^{-1}\alpha_{2}^{*}
\\ 
-P\beta_{1}^{*}\beta_2\alpha_1+P\beta_{3}^{*}\alpha_2&&\beta_2&&-P\beta_1^{*}\beta_{2}\alpha_{2}^{*}-P\beta_{3}^{*}\alpha_{1}^{*}
 \\-P\beta_{1}^{*}\beta_3\alpha_1-P\beta_{2}^{*}\alpha_2 
&&\beta_3&&-P\beta_1^{*}\beta_{3}\alpha_{2}^{*}+P\beta_{2}^{*}\alpha_{1}^{*} 
\end{array}
\right)
\end{equation} 
 yields the parametrization due to Anselm et al [7]:
\begin{equation}
 \left(\begin{array}{ccc} C_{12}C_{13} &-S_{12}&-C_{12}S_{13}\\ 
S_{12} C_{13}C_{23}e^{-i\alpha}-S_{13}S_{23}&
C_{12} C_{23}e^{-i\alpha} & -S_{12} S_{13}C_{23}e^{-i\alpha}-
C_{13}S_{23}\\
S_{12}C_{13}S_{23}+S_{13}C_{23}e^{i\alpha}&C_{12} S_{23}&
-S_{12} S_{13}S_{23}+ C_{13}C_{23}e^{i\alpha} 
\end{array}\right)
\end{equation}

\section{SU(4) and three Kobayashi-Maskawa phases}
We now consider the $N=4$ case. Let $\bg$ denote a four dimensional 
complex unit vector. Then, for $|\gamma_1|^2 +|\gamma_2|^2 <1$,
\begin{equation}
C(\bg)  = \left(\begin{array}{cccc} Q^{-1}&0&0&\gamma_1\\ 
-Q\gamma_{1}^{*}\gamma_{2}&QR^{-1}&0&\gamma_2\\
-Q\gamma_{1}^{*}\gamma_{3}&-QR\gamma_{2}^{*}\gamma_{3}&R\gamma_{4}^{*}
&\gamma_3\\
-Q\gamma_{1}^{*}\gamma_{4}&-QR\gamma_{2}^{*} \gamma_{4}&-R\gamma_{3}^{*}
&\gamma_4
\label{13}
\end{array}
\right)
\end{equation}
where $Q=(1-|\gamma_1|^2)^{-1/2}$, $R=(1-|\gamma_1|^2-|\gamma_2|^2)
^{-1/2}$, is in $SU(4)$. The unit vector $\bg$ is a label for right $SU(3)$ 
cosets in $SU(4)$ and $C(\bg)$ 
is a coset representative. So, except on a subset of measure zero, for a 
$C\in SU(4), |C_{14}|^{2} + |C_{24}|^{2}<1$, there is a unique sequence of 
complex unit vectors $\bg,\bb,\ba$ of dimensions $4,3,2$ respectively, such 
that  
\begin{equation}
C=C(\bg,\bb,\ba)= C(\bg)
\cdot \left(\begin{array}{cc} B(\bb,\ba)&0\\ 
0&1\end{array}
\right)
\label{14}
\end{equation}
Now we multiply $C$ on the left and right by independent diagonal $SU(4)$
matrices, get the transformation laws for $\bg,\bb,\ba$, and then construct 
the invariants. 
\begin{equation}
C^\prime = D(\theta^\prime)~ C(\bg,\bb,\ba)~ D(\theta)= C(\bg^\prime,\bb
^\prime,\ba^\prime)
\end{equation}
where $D(\theta)=diag(e^{i\theta_1+i\theta_2+i\theta_3}, e^{-i\theta_1+
i\theta_2+i\theta_3},
e^{-2i\theta_2+i\theta_3}, e^{-3i\theta_3})$ and 
$D(\theta^\prime)$ is defined similarly. For simplicity, let $B(\bb,\ba)$ also 
denote the $4\times4$ matrix obtained by an appropriate bordering. Then, 
because of the way we parametrized $D(\theta)$ and $D(\theta^\prime)$, we find 
\begin{eqnarray}
C'&=&C(\bg^\prime)B(\bb^\prime,\ba^\prime)\nonumber\\
&=& D(\theta^\prime)C(\bg)B(\bb,\ba)D(\theta)\nonumber \\
&=&D(\theta^\prime) C(\bg)~diag(e^{i\theta_3},e^{i\theta_3},e^{i\theta_3},
e^{-3i\theta_3})\nonumber\\
&&\times B(\bb,\ba)~diag(e^{i\theta_1+i\theta_2},e^{-i\theta_1+
i\theta_2},e^{-2i\theta_2},1) 
\end{eqnarray}
The expressions for $\bg^\prime$ are easy to read off
\begin{eqnarray}
\gamma_{1}^{\prime} &=& \gamma_{1}e^{i(\theta_{1}^\prime+
\theta_{2}^{\prime}+\theta_{3}^{\prime}-3\theta_3)}~~~;~~~
\gamma_{2}^{\prime} = \gamma_{2}e^{i(-\theta_{1}^\prime+
\theta_{2}^{\prime}+\theta_{3}^{\prime}-3\theta_3)}\nonumber\\
\gamma_{3}^{\prime}& =& \gamma_{3}e^{i(-2\theta_{2}^\prime+\theta_{3}^\prime
-3\theta_2)}~;~
\gamma_{4}^{\prime} = \gamma_{4}e^{-3i(\theta_{3}^\prime+\theta_{3})}
\label{9}
\end{eqnarray} 
A little algebra shows that
\begin{eqnarray}
&&D(\theta^\prime) C(\bg)~diag(e^{i\theta_3},e^{i\theta_3},e^{i\theta_3},
e^{-3i\theta_3})\nonumber \\&=&C(\bg^\prime)~diag(e^{i(\theta_{1}^\prime+
\theta_{2}^{\prime}+\theta_{3}^{\prime}+\theta_3)},
e^{i(-\theta_{1}^\prime+
\theta_{2}^{\prime}+\theta_{3}^{\prime}+\theta_3)},
e^{-2i(\theta_{2}^\prime+
\theta_{3}^{\prime}+\theta_{3})},1)
\end{eqnarray}
so that the rest reduces to an $SU(3)$ problem in $3\times 3$ matrix form
\begin{eqnarray}
& &B(\bb^\prime,\ba^\prime)= \nonumber \\
& &diag(e^{i(\theta_{1}^\prime+
\theta_{2}^{\prime}+\theta_{3}^{\prime}+\theta_3)},
e^{i(-\theta_{1}^\prime+
\theta_{2}^{\prime}+\theta_{3}^{\prime}+\theta_3)},
e^{-2i(\theta_{2}^\prime+
\theta_{3}^{\prime}+\theta_{3})})\nonumber\\ 
&& \times B(\bb,\ba)
diag(e^{i\theta_1+i\theta_2},e^{-i\theta_1+
i\theta_2},e^{-2i\theta_2})
\end{eqnarray}
which is just the same as in $(\ref{6})$ with the replacements $
\theta_1\rightarrow \theta_1, \theta_2\rightarrow \theta_2,
\theta_{1}^\prime \rightarrow \theta_{1}^{\prime}, 
\theta_{2}^\prime \rightarrow 
\theta_{2}^\prime+\theta_{3}^\prime +\theta_{3}$. Making these changes in 
$(\ref{7})$ and $(\ref{8})$ we see that for the $SU(4)$ problem to accompany 
$(\ref{9})$, we have,
\begin{eqnarray}
\alpha_{1}^{\prime} &=& \alpha_{1}e^{i(\theta_{1}^\prime+
\theta_{2}^{\prime}+\theta_{3}^{\prime}-\theta_{1}+\theta_2+\theta_3)}~~~;~~~
\alpha_{2}^{\prime} = \alpha_{2}e^{-i(\theta_{1}^\prime+
\theta_{2}^{\prime}+\theta_{3}^{\prime}+\theta_{1}+\theta_2+\theta_3)}
\label{10}\\
\beta_{1}^{\prime}& =& \beta_{1}e^{i(\theta_{1}^\prime+\theta_{2}^\prime
+\theta_{3}^{\prime}-2\theta_2+\theta_3)}~;~
\beta_{2}^{\prime} = \beta_{2}e^{i(-\theta_{1}^{\prime}+\theta_{2}^{\prime}
+\theta_{3}^{\prime}-2\theta_2 +\theta_3)}\nonumber\\
\beta_{3}^{\prime} &=& \beta_{3}e^{-2i(\theta_{2}^{\prime}+
\theta_{3}^{\prime}+\theta_{2}+\theta_3)}
\label{11}
\end{eqnarray} 
From $(\ref{9})$, $(\ref{10})$ and $(\ref{11})$ we need to construct the 
invariants. The six `Cabibbo' angles may be taken to be given by
$|\alpha_1|,|\beta_1|,|\beta_2|,|\gamma_1|,|\gamma_2|,|\gamma_3|$. The three 
KM phases can be obtained  systematically as follows. Since $\theta_1$ 
is involved only in $\ba^{\prime}$ , not in $\bb^\prime$ and $\gamma^\prime$, 
we see that the $\alpha$'s must enter only in the form 
$\alpha_1 \alpha_{2}^{*}$ 
which obeys 
\begin{equation}
\alpha_1 \alpha_{2}^{*}\rightarrow \alpha_1 \alpha_{2}^{*}
e^{2i(\theta_{1}^\prime+\theta_{2}^{\prime}+\theta_{3}^{\prime}+\theta_{2}
+\theta_3)}
\end{equation}
Next, we see that $\theta_2$ is not involved in the $\gamma^\prime$'s at all, 
so we form independent expressions in $\alpha_1 \alpha_{2}^{*}$ and the 
$\beta$'s 
in which $\theta_2$ goes away. 
\begin{eqnarray}
\alpha_1 \alpha_{2}^{*}\beta_1 &&\rightarrow \alpha_1 \alpha_{2}^{*}\beta_1
e^{3i(\theta_{1}^\prime+\theta_{2}^{\prime}+\theta_{3}^{\prime}+\theta_{3})}
\nonumber\\
\alpha_1 \alpha_{2}^{*}\beta_2 &&\rightarrow \alpha_1 \alpha_{2}^{*}\beta_2
e^{i\theta_{1}^{\prime}+3i(\theta_{2}^{\prime}+\theta_{3}^{\prime}+\theta_{3})}
\nonumber\\
\alpha_1 \alpha_{2}^{*}\beta_3 &&\rightarrow \alpha_1 \alpha_{2}^{*}\beta_3
e^{2i\theta_{1}^\prime}
\end{eqnarray}
These three quantities and the four $\gamma$'s involve 
$\theta_{1}^\prime,\theta_{2}^{\prime},\theta_{3}^{\prime},\theta_{3}$. We 
now form independent combinations in which $\theta_3$ drops out. They are
\begin{eqnarray} 
\gamma_1 \gamma_{2}^{*}&& \rightarrow
\gamma_1 \gamma_{2}^{*}e^{2i\theta_{1}^{\prime}}\nonumber\\
\gamma_1 \gamma_{3}^{*}&& \rightarrow
\gamma_1 \gamma_{3}^{*}e^{i(\theta_{1}^{\prime}+3\theta_{2}^{\prime})}
\nonumber\\
\gamma_1 \gamma_{4}^{*}&& \rightarrow
\gamma_1 \gamma_{4}^{*}e^{i(\theta_{1}^{\prime}+\theta_{2}^{\prime}+
4\theta_{3}^{\prime})}\nonumber\\
\alpha_1 \alpha_{2}^{*}\beta_1 \gamma_4 &&\rightarrow \alpha_1 \alpha_{2}^{*}
\beta_1\gamma_4
e^{3i(\theta_{1}^{\prime}+\theta_{2}^{\prime})}\nonumber
\\
\alpha_1 \alpha_{2}^{*}\beta_2\gamma_4 &&\rightarrow \alpha_1 \alpha_{2}^{*}
\beta_2\gamma_4
e^{i(\theta_{1}^{\prime}+3\theta_{2}^{\prime})}\nonumber
\\
\alpha_1 \alpha_{2}^{*}\beta_3 &&\rightarrow \alpha_1 \alpha_{2}^{*}\beta_3
e^{2i\theta_{1}^{\prime}}
\end{eqnarray}
Here $\theta_{3}^{\prime}$ appears only in the rule for 
$\gamma_1\gamma_{4}^{*}$ so we just drop it. Then we quickly find a choice 
of three independent invariants:
\begin{equation}
arg(\alpha_1 \alpha_{2}^{*}\beta_{1}^{*}\beta_2\beta_3)~~;~~
arg(\beta_{1}\beta_{2}^{*}\gamma_{1}^{*}\gamma_2)~~;~~
 arg(\beta_2 \beta_{3}^{*}\gamma_{2}^{*}\gamma_3\gamma_4)~~;~~
\end{equation}
The first of the three $SU(4)$ invariants is the same as the single $SU(3)$ 
invariant. This is explained by the observation that after the 
$\gamma^\prime$'s 
were determined in $(\ref{9})$, the determination of the $\beta^\prime$'s and 
$\alpha^\prime$'s was reduced to the $SU(3)$ level problem - the $SU(4)$ 
expressions 
for the $\beta^\prime$'s and $\alpha^\prime$'s arise from those for $SU(3)$ in $(\ref{7})$
and $(\ref{8})$ by the replacements   
$\theta_1\rightarrow \theta_1, \theta_2\rightarrow \theta_2,
\theta_{1}^\prime \rightarrow \theta_{1}^\prime, \theta_{2}^\prime \rightarrow 
\theta_{2}^\prime+\theta_{3}^\prime +\theta_{3}$.

The recursive procedure given above can easily be extended to $N$ generations. 

\section{Comparison with some existing parametrizations of the mixing matrix 
for N=4}
The parametrization due to Barger et al [8] and Oakes [9] correponds to 
choosing $\beta_2, \gamma_2, \gamma_3$ complex and all others real. 
Thus, on putting
\begin{eqnarray}
\gamma_1 &=& C_1~~;~~\gamma_2 = -S_1 C_2 e^{-i(\delta_1 +\delta_3)}
~~;~~\gamma_3 = -S_1 S_2 C_4 e^{-i\delta_2}~~;~~\gamma_4=-S_1 S_2 S_4\nonumber
\\
\beta_1 &=& C_3~~;~~\beta_2 = S_3 C_6 e^{-i\delta_3}~~;~~\beta_3 =S_3 S_6
\nonumber\\
\alpha_1 &=& C_5~~;~~\alpha_2 = S_5 
\end{eqnarray} 
in $(\ref{14})$ and interchanging the first and the fourth columns and 
the second and the third we recover the parametrization in [8] and [9]
after multiplying the second and the third row by phase factors 
$e^{i(\delta_1 +\delta_3)}$ and $e^{i\delta_2}$ respectively.

The parametrization of the mixing matrix for $N=4$ due to Anselm et al 
[7] is less economical. It corresponds to distributing the three phases  
four quantitities $\beta_3,\gamma_2, \gamma_3, \gamma_4 $ with all others real:
\begin{eqnarray}
\gamma_1 &=& -S_{12}~;~
\gamma_2 = C_{12} C_{23}C_{24} e^{-i\alpha}
~;~\gamma_3 = C_{12} (S_{23} C_{24} C_{34}-S_{24}S_{34})e^{i\gamma}~
;~\gamma_4=C_{12}(S_{23}C_{24}S_{34}e^{-i(\beta+\gamma)}+S_{24}C_{34})
e^{-i\beta}\nonumber
\\
\beta_1 &=& -S_{13}~;~\beta_2 = C_{13}S_{23}/\sqrt{(1-C_{23}^{2}C_{24}^{2})}
 ~;~\beta_3=C_{13}C_{23}S_{24}e^{i(\alpha-\beta)}
/\sqrt{(1-C_{23}^{2}C_{24}^{2})}
\nonumber\\
\alpha_1 &=& C_{14}~;~\alpha_2 = -S_{14} 
\end{eqnarray}
 
Substituting these expressions in $(\ref{14})$ one obtains the results 
of Anselm et al after suitable permutation of the columns and multiplication  
of second and fourth row by  factors $e^{i\alpha}$ and $e^{i(\beta+\gamma)}$. 

The parametrization due to Harari and Leurer [14] corresponds to choosing 
$\beta_1,\gamma_1,\gamma_2$ complex with all others real. Thus on putting 
\begin{equation}      
\alpha_1= S_{12}, \alpha_2= C_{12}, \beta_1 = S_{13}e^{-i\delta}, \beta_2 = 
S_{23} C_{13} , \beta_3 = C_{23} C_{13} 
\end{equation}
and 
\begin{equation}
\gamma_1=S_{14}e^{-i\delta_{14}}~;~\gamma_2=C_{14} S_{24}e^{-i\delta_{24}}~
\gamma_3=C_{14}C_{24}S_{34}~;~\gamma_4=C_{14}C_{24}C_{34}
\end{equation}
in $(\ref{14})$ we recover their results.
In fact in the Harari-Leurer parametrization, to go from $3~(or~N-1)$ 
generations to $4~(or~N)$ generations, one needs to multiply the 
(appropriately 
augmented) mixing matrix at $3~(or~ N-1)$ generations, on the left, by a 
matrix 
consisting of $3~(or~N-1)$ factors. In the present case, it is easily seen 
that 
the three matrices when multiplied out have precisely the same structure as in 
$(\ref{13})$. In general, the $N-1$ factors when multiplied out 
precisely correspond to the coset representative of $SU(N)/SU(N-1)$  
characterized by an $N$-dimensional complex unit vector with its first 
$N-2$ components complex and the rest real.

\section{Phases in the mixing matrix and the Bargmann invariants}
It is known that, under rephasing, apart from the obvious  invariants 
$|V_{\alpha i}|$, the magnitudes of the matrix elements of the mixing matrix, 
the following quantities, quartic in $V$'s,
\begin{equation}
t_{ \alpha i \beta j}\equiv V_{\alpha i} 
V_{\beta j}V_{\alpha j}^{*}V_{\beta i}^{*}
\end{equation}
are invariant under the rephasing transformations
\begin{equation}
V_{\alpha i}\rightarrow  e^{i\theta_{\alpha}^{\prime}}V_{\alpha i}
e^{i\theta_{i}}
\end{equation}
It is also evident that this  set of invariants remains unchanged under 
row and column permutations.
 
In the present context, these invariants were first discussed by 
Jarlskog [15] and by Greenberg [16] for the case of three generations  (for 
which there is only one independent invariant) and were later 
generalized to N-generations by Nieves and Pal [17]  who showed that of these  
the following $(N-1)(N-2)/2$ quantities can be taken as independent 
\begin{equation}
 t_{\alpha i 1 N}~~;~~\alpha \leq i, \alpha \neq 1,
i\neq N
\end{equation} 
It can easily be verified by explicit calculations that the invariant phases 
given earlier for $N=3,4$ precisely coincide with 
$ arg (t_{\alpha i 1 N})~~;~~\alpha \leq i, \alpha \neq 1,$. 

We would like to bring out the connection between these and the 
Bargmann invariants introduced by Bargmann in the context of  Wigner's 
unitary-antiunitary theorem. If $\psi_1,\psi_2\,\cdots,\psi_n$ are any $n$ 
vectors in a Hilbert space, with no two consecutive ones being orthogonal, 
the n-vertex Bargmann invariant is 
\begin{equation}
\Delta_n (\psi_1,\psi_2,\cdots,\psi_n)= <\psi_1|\psi_2><\psi_2|\psi_3>
\cdots <\psi_n|\psi_1>
\end{equation} 
It is easily seen that, under a common unitary transformation applied to 
all the 
$\psi$'s, and also, under independent phase changes of the $\psi$'s,
$\Delta_n$ remains unchanged. As an aside, we would like to remark here 
that there exists a deep connection between Bargmann invariants 
and the geometric phase as has been lucidly brought out by  
Mukunda and Simon [19].  
 
To see the relevance of Bargmann invariants in the present context, 
notice that $V$ being a unitary matrix can be thought of as effecting a 
change of basis from one set of orthonormal vectors $|f_i>$ to another 
$|e_\alpha>$ so that $V_{\alpha i}= <f_i|e_\alpha>$ and one can express 
$V_{\alpha i}V_{\beta j}V_{\alpha j}^{*}V_{\beta i}^{*}$ as a Bargmann 
invariant
\begin{equation}
V_{\alpha i}V_{\beta j}V_{\alpha j}^{*}V_{\beta i}^{*}  
= <e_\alpha|f_i><f_i|e_{\beta}><e_{\beta}|f_j><f_j|e_\alpha>
\end{equation}

\section{Summary}
To summarize, the parametrization proposed here has the following special 
features:
\begin{itemize}

\item Introduction of $N^{th}$ generation requires one new $N\times N$ matrix 
determined by one $N$-dimensional complex unit vector, a $SU(N)/SU(N-1)$ 
coset representative,  multiplying the 
complete matrix at previous generation level after augmenting its dimension 
by one through bordering the last column and row suitably. 

\item All the invariants for $N-1$ generations remain invariants for 
$N$-generations as well.
 
\item One matrix of ours determined by an $N$-dimensional unit vector 
corresponds to a product of $N-1$ factors of Harari and Leurer.

\item The existing parametrizations are easily read off from our general 
expressions. 

\item Opens up new possibilities for alternatives parametrizations 
which may be phenomenologically useful, particularly for $N\geq4$.

\item The connection between the rephasing invariants and the Bargmann 
invariants is brought out.
\end{itemize}

We hope that the unified approach to parametrization of the mixing matrix 
developed here will prove to be phenomenologically useful as well. 
In particular, the connection between the phases and the Bargmann invariants 
brought out here may provide a new perspective on their origin. 

\vskip1cm

One of us (SC) is grateful to Prof V. Gupta for asking a question which 
initiated this work. We are also grateful to Prof R. Simon for numerous 
discussions.

\end{document}